\documentclass{camera}
\usepackage{graphicx,bm,color}
\usepackage{subfigure}
\def \bit{\begin{itemize}}
\def \eit{\end{itemize}}
\def \ben{\begin{enumerate}}
\def \een{\end{enumerate}}
\def \bsubeq{\begin{subequations}}
\def \esubeq{\end{subequations}}
\def \beq{\begin{equation}}
\def \eeq{\end{equation}}
\def \beqa{\begin{eqnarray}}
\def \eeqa{\end{eqnarray}}
\begin{document}
 
\title{Quark stars: their influence on Astroparticle Physics}
\author{Sanjay K. Ghosh}
\organization{Department of Physics, Bose Institute, 
          93/1, A.P.C. Road, \\ Kolkata 700 009, India. \\
and \\
Centre for Astroparticle Physics \& Space Science, Bose Institute, Block EN, Sector V, Salt Lake, Kolkata - 700 091, India}

\maketitle

\begin{abstract}
We discuss some of the recent developments in the quark star physics along with the consequences of possible hadron to 
quark phase transition at high density scenario of neutron stars and their implications on the Astroparticle Physics.
\end{abstract}

\section{Introduction}

The underlying quark structure of hadrons suggets the possibility of a quark-hadron phase transition at extreme 
conditions of high tempereature and/or density. Since a compact object like neutron star (NS) provide the 
natural scenario of high density, the suggestion for the existence of quark core inside such massive compact objects was put forward by
Ivanenko and Kurdgelaidze (1969). The existence of 3-flavour quark star or strange star (QS), made up
of $u$, $d$ and $s$ quarks, was suggested by Itoh (1970). In general, the two flavour matter can not be more stable compared to nucleonic matter. The presence of $s$ quarks, along 
with $u$ and $d$ quarks, provides
an additional fermi well which would result in the lowering of the energy of the 3- flavour quark matter or strange quark matter (SQM) compared to 2- flavour quark matter. Since $s$ quark has larger mass, the situation would be more favourable at higher densities. Such possibilities were recognized by Bodmer (1971)
and led Witten (1984) to conjecture that SQM may be the true ground state of strongly interacting matter at
high densities. Since then, this field has become an active area of research. The possibilty of a quark -hadron phase
transition in the early universe (high temperature and small chemical potential) scenario and attempts to 
produce such a matter in the laboratory through heavy ion collision reactions (Alam, Sinha \& Raha, 1996) using 
accelerators have made 
this field more interesting as well as intriguing. 

One of the major difficulty in the theoretical calculations in these areas of research is 
the fact that the Quantum Chrmodynamics (QCD) perturbative series shows poor 
convergence except for very small
coupling at very high temperatures ($ \alpha_s < 0.5$, $T \sim 10^{5} T_c$). The perturbative treatment
at high density also fails. The lattice calculations are still not reliable 
in the high density regime applicable to
NS. Hence the high density systems are studied using the QCD inspired phenomenological models. 
Numerous model
calculations have predicted a stable quark matter system within finite parameter 
ranges (Ghosh \& Sahu, 1993). 

Presently, the major technical advancements in both ground as well as satellite based
observations are producing huge amount of data. There exist large number of observational data on 
mass-radii of NS. But, except very soft equations of state (EOS) most of the other EOS can
explain these static properties within the error bars (Schaffner-Bielich, 2007). So it is necessary
to study different dynamical features of quark matter in its various forms so as to predict
unambiguous signatures of QS and validate them using the available observational data.

In this article, we will review the present status of theoretical understandings of the role of quark matter in 
the development of astroparticle physics, more specifically, the physics of compact objets with quark core 
(hybrid stars or HS)
and QS in the light of various observational studies along with the efforts for the search of strange matter in cosmic rays. 
We divide the article in few sections. Section 2 deals with symmetry structures of quark matter at
high densities. Compact stars' pulsating modes and their importance is discussed in section 3.
QCD phase transition and its consequences are discussed in section 4. Section 5 deals with Strangelets.
Summary and discussions are given at the end.

\section{Symmetries and different quark matter phases}
At very high densities and small temperatures, interaction is very weak because of asymptotic freedom. 
Fermi energy being almost same as chemical potential, adding or subtracting a particle 
does not cost much. Presence of attractive potential allows the addition of a pair of particles with the 
attractive channel quantum number resulting into an energetically more favourable configuration.
For QCD, the colour coulomb interaction is attractive between quarks having
antisymmatric colour wave function. So the pairing in colour  space is
really a natural consequence of QCD theory itself. Since pair of quarks can not be colour
singlet, the cooper pair condensation in colour space will spontaneously break the color symmetry
and  gluons will acquire mass (Rajagopal \& Wilczek, 2000; Alford {\it et al.}, 2007).

Again since, quarks also have flavour and spin along with colours, there
will be many variety of patterns due to quark cooper pairing. For example, with increasing density, 
fermi surfaces of $u$, $d$ and $s$ will come closer and one will get first unpaired,
then 2-flavour ($u$ \& $d$) paired and finally 3-flvaour paired (Colour flavour locked or CFL) quark matter.
(Casalbuoni, 2004; Rajagopal \& Wilczek, 2000). 

Depending on the value of superconducting gap,
various form of pairing is expected to occur inside compact stars (Alford {\it et al.}, 2007) with varied consequences.
For example, the cooling rate is given by direct URCA
and goes
as $T^6$ for unpaired quark matter, nuclear matter at higher densities and some
other forms of matter (Iwamoto, 1982; Ghosh, Phatak \& Sahu, 1994, 1996). At very low densities, for proton
fraction lower than 0.1, modified URCA is the relevant process and
cooling rate varies as $T^8$. 
In case of CFL matter, all the fermions are gapped and the emissivity due to Goldstone modes are
suppressed by the Boltzman factor. 
So the emissivity from CFL matter is governed by the processes involving $\phi$, the massless Goldstone
boson associated with breaking of $U(1)_B$. 
Neutrino emission from such processes  goes as $T^{15}$ (Jaikumar, Prakash \& T. Sch\"{a}fer, 2002;
Reddy, Sadzikowski \& Tachibana, 2003).
The present observational scenario indicates that some NS cool
much faster than others. So, most probably, lighter NS 
cool following the modified URCA process whereas the heavier NS
contain some form of matter, other than CFL, which follows direct URCA process (Alford {\it et al.}, 2007). 

\section{Pulsating modes and Gravitation Wave}
In rotating 
neutron stars (RNS), presence of different restoring forces ( pressure, gravity, coriolis force etc.)
give rise to various modes of oscillations. There are also various damping mechanisms such as viscosity, 
electromagnetic \& gravitational radiation, neutrino emission etc.

In the case of radial oscillation (RO), the star oscillates around its equilibrium configuration maintaining its shape.
For non-radial oscillation (NRO), shape of the star is not preserved. The RO 
may be regarded as a special case of NRO with angular momentum quantum number $l = 0$.

RO in compact stars have been studied by many authors (Benvenuto \& Horvath, 1991;
Gondek \& Jdunik, 1999; Kokkotas \& Ruoff, 2001). 
Different observations, such as radio subpulses of pulsars, short duration spikes in many pulsars and 
discovery of submillisecond oscillation in celestial X-ray source, have motivated the study of oscillation 
in compact stars (Boriakoff, 1978; Cordes, 1976; Van-Horn, 1980).

The observation or non-observation of pulsation may also help us to distinguish between the NS and QS. 
Vath \& Chammugam (1992) showed that the periods of RO of QS behave very differently from NS.
Instead of having smallest possible period for a given mode, the period of all modes for QS go to zero when the 
central density of the QS approaches its smallest possible value. Furthermore, bulk viscosity being larger for 
normal quark matter, RO for QS would damp faster than NS.
For CFL QS, damping will be much lesser compared to both NS and normal QS.

Recent studies indicate that the NRO, {\it i.e.} $ l > 0$ oscillations may be more promising
compared to the RO for distinguishing between the NSs with different internal composition. 
This is more so as the RO do not couple to gravitational radiation.

In general, for any star there are two types of NRO, namely spheroidal or polar and toroidal or axial perturbation 
mode (Kokkotas \& Schmidt, 1999). For a Newtonian non-rotating
perfect fluid star, all the modes are spheroidal. The axial modes are the trivial modes 
with zero frequency and without any variation of pressure and density. But for relativistic case, though the 
picture is otherwise similar, the dynamic space time gives rise to gravitational w- mode. The main characteristics
of the  w - mode is the rapid damping of the oscillation which increases as the compactness of the star decreases.
Moreover, the w- modes do not induce any significant fluid motion (McDermott, Van Horn \& Hansen, 1988).

When the star starts rotating, the trivial axial modes become non degenerate and a new family emerges; the r-
modes which are analogous to Rossby waves in earth's ocean. The r- mode pulsation in compact stars are unique in 
the sense that they are unstable due to the emission of gravitational radiation at all rates of rotation
(Andersson \& Kokkotas, 2001). The restoring force of these r- modes is the coriolis force and
it transfers the star's angular momentum into gravitational radiation. Since several large scale 
gravitational wave detectors are operational, 
the study of NRO, especially, r-mode pulsation which are
unique characteritics of rotating star, has become important (Andersson, Kokkotas \& Sterigioulas, 1999). 

For a RNS, there is a critical frequency above which
the r-mode instability sets in, angular momentum gets transfered
to gravitational wave and star spins down (Andersson \& Kokkotas, 2001).
This r-mode instability is limited by the viscous damping. For a larger viscosity
the critical spin at which r-mode becomes unstable is higher. Since the bulk viscosity of 
normal SQM is larger than that of normal NS matter, 
an observation of newly born pulsar spinning near the Keplarian limit would provide
the evidence for a QS (Madsen, 2000). Similar situation is expected
for 2SC quark star as well. CFL matter has very small shear damping and bulk viscosity. 
So the heating effect due to the viscous dissipation is less in CFL stars and r-mode 
damping becomes more important for their evolution. So, except for first few hundred years, 
CFL stars will cool very slowly and can exist at higher temperatures for many years (Zheng, Yu \& Li, 2006).

The unstable r-mode seems to affect the QS differently as compared to 
NS (Andersson, Jones \& Kokkotas, 2002). Unlike NS, the onset of r-mode instability, instead of leading 
to the thermo-gravitational runaway, results in the evolution of QS to a quasiequilibrium state.
Moreover, for QS, r-mode instability never grow to large amplitudes.

The gravitational wave bursts induced by r-mode spin down of
HS has also been suggested. The continuous
emission of gravitational waves due to r-mode instability from a star can
induce a sudden variation in its structure and composition generating further
bursts of gravitational waves. This scenario is more probable for HS due to the
surface tension between the hadronic and quark matter (Drago, Pagliara \& Berezhiani, 2006). 

\section{QCD Phase transition and Compact Objects}

The r-mode instability and the corresponding spin down will cause an increase in
the central density of the star. This sudden increase may trigger a
phase transition inside the core of NS.
In SQM, the strangeness fraction, {\it i.e.} the ratio of
strange quark and baryon number densities, will be unity if one considers
u, d and s masses to be same. Even for relativistic quark masses
($m_s > m_u \sim m_d$), the strangeness fraction at high density is not much
smaller than unity. On the contrary, the strangeness fraction in hadronic
matter is usually small. Even with hyperons, the strangeness fraction is
smaller compared to quark phase. Off course with kaon condensation,
the situation may be different. But then, with a kaon condensation
inside NS, the transition to quark matter is found to
be pushed towards much higher densities (Bhattacharyya {\it et. al.}, 1997). So the transition
from hadronic to quark matter may be associated with large strangeness production.
Initially hadronic matter that gets deconfined to 2-flavour matter consist predominantly 
of u \& d and some s quark due to the hyperon
population. The matter is certainly out of equlibrium. The weak interaction converts
this chemically non-equilibrium matter to equilibrated matter with roughly equal
number of u, d and s quarks. This conversion is associated with the release of large
amount of energy in the form of neutrinos with average energy $\sim 100$ Mev (Ghosh, Phatak \& Sahu, 1996).
The total amount of energy released is in agreement
with the $\gamma$- ray burst energy.

Conversion of neutron matter to SQM may be treated as a two-step process.
Deconfinement of nuclear matter, which is
predominantly n, p, e$^{-}$ matter, to u , d and e$^{-}$ matter takes place in the first step in a strong interaction
time scale. The final composition of the 2-flavour quark matter
may be determined
from the nuclear matter EOS by enforcing the baryon number conservation
during the conversion process. This can be studied as the evolution of combustion front moving outward in the
radial direction inside the model NS in the special relativistic formalism
 
In the second step, this 2-flavour quark matter gets converted to SQM.
The strange quarks are generated from the excess of down quarks via the non-leptonic and semileptonic weak process,
resulting into a charge neutral $\beta$ equilibrated SQM. Here again, one may assume the existence of a 
conversion front in the core of
the star that propagates radially outward leaving behind the SQM as
the combustion product. The time taken to convert the whole star in the above two processes are about $10^{-3}$ and
$100$ seconds respectively (Bhattacharyya {\it et al.}, 2006).

For NS environment, one should really study the conversion along with 
general relativistic (GR) effect. The GR effect gives rise to different conversion fronts propagating
with different velocities along
different radial directions (Bhattacharyya {\it et al.}, 2007a). 

Hadronic matter to quark matter phase transition and the resulting neutrino emission can be
studied for a RNS. Here, the emission gets predominantly confined to a small angle 
provided the core of the star is in a mixed phase and the size of this mixed phase is small.
This neutrino beaming may be the missing link that causes the GRB. In general,
cross section for the reaction \( \nu + \bar\nu \rightarrow e^- + e^+ \) is very small. 
The effect of rotation, along with the general relativity, enhances the energy deposition rate substantially and 
can provide a very efficient engine for the gamma ray bursts (Bhattacharyya {\it et al.} 2007b).

\section{Magnetar and quark stars}

The highly periodic pulsation of pulsars are attributed to strong magnetic field ($10^{8} - 10^{12}$G)
(Michel, 1982, Lugones, 2005).
The study of the role of magnetic field has
become more important with the discovery of magnetars. Observed as Soft Gamma Repeaters
(SGR) and Anamolous X- ray Pulsars (AXP), these sources are believed to be directly powered by the decay of 
ultrastrong $\sim 10^{14} - 10^{15}$G magnetic field and hence are known collectively as magnetars.

There are two reasons which makes QS a possible candidate for magnetar. They are (a) flares with 
luminosities from some of the SGRs far exceeding the Eddington limit 
$L_{Edd} \sim 1.5 \times 10^{38} (\frac{M}{M_{\odot}}) erg/s$ and 
(b) the inferred presence of ultra-strong magnetic field.

{\bf{Case (a)}}  Eddington limit is the critical luminosity of a normal  star when the photon radiation pressure 
from the surface of the star becomes equal to the inward gravitational pressure. As the SQM is bound via strong interaction 
rather than gravity, QS is not subject to Eddington limit and can radiate at the luminosities greatly 
exceeding Eddington limit (Usov, 1998, 2001). The giant bursts of SGR0526-66 
and SGR1900+14 can be explained in a model where the burst radiation is produced from the bare quark surface of the QS 
heated by the impact of a massive comet like object (Zhang, Xu, \& Qiao 2000).

{\bf {Case (b)}}  In general for pulsars the loss of rotational energy may account for all the observed radiation.
But, for magnetars, the energy emitted in both the quiescent emission (both SGRs and AXPs) and flares (SGR) far
exceed the loss in their rotational energy over the same period. The only known source of energy for these emissions is the magnetic energy. The difference between the nature of the normal pulsars and those of SGRs and AXPs raises the possibility
that magnetars may have different internal states as compared to ordinary pulsars. One of the possible scenario may be the
formation of strongly magnetized core by the strong interaction during the collapse of the progenitor 
(Soni and Bhattacharya, 2004).
Such a highly magnetized core is possible due to the phase transition from neutron matter to quark matter, 
neutral pion condensate being the ground state of such matter
at 5-6 times nuclear matter densities (Soni \& Bhattacharya, 2004; kutschera, Broniowski \& Kotlorz, 1990).
Such a state carries a large magnetic moment and hence
can give rise to a strongly magnetized core.

\section{Strangelets and detection : cosmic ray search}
While only SQM with very large baryon numbers was initially thought to
be favorable (in terms of stability), later calculations have shown (Farhi \& Jaffe, 1984; Mustafa \& Ansari, 1995,1996)
 that small lumps of SQM can also be stable.
The occurrence of stable (or metastable) lumps of SQM or strangelets would lead to
many rich consequences (Madsen, 1999). In general,
the stability of strangelets with atomic number A up to 20 - 30 depend rather sensitively
on the parameter values (like the Bag constant) and an underlying shell-like structure. For larger strangelets
(A $>$ 40), the stability appears to be more robust (Madsen, 1994). A discerning property
of such strangelets would be an unusual charge to mass ratio ($\frac{Z}{A} \ll 1$) (Banerjee {\it et al.}, 1999).
The obvious place to look for such strangelets would be in the cosmic ray flux. 

The detection of strangelet would depend on their mechanism of propagation through the 
terrestrial atmosphere. It has been shown that
a small strangelet with $A \sim 100$ and $Z \sim 2$ can propagate through the atmosphere and reach the mountain 
altitudes (3-4 Km) with 
the modified $A \sim 300 - 400$ and $Z \sim 10 - 20$ and a small $\beta$ (Banerjee {\it et al.}, 2000a, 2000b). 

Several experiments, both ground (Ambrosio, 2000, Balestra, 2006, Cecchini, 2008) and satellite (Madsen, 2005) based, have been designed for the detection of strangelets. Here, one should note that to investigate rare events such as strangelets in cosmic ray, not only a large array of detector is needed, the choice of detector is also extremely important. One of the suitable choice is polymer or plastic detector. But the commonly used plastic detectors such as CR39 and Lexan polycarbonate have detection threshold of
$\frac{Z}{\beta} \sim 6 - 10$ and $57$ respectively. On the other hand, if the strangelets accumulate large charge
during propagation, their $\frac{Z}{\beta}$ would be much larger. In such a situation CR39 like detectors would register a huge low - Z noise which would make the detection of rare events extremely difficult.

In contrast, recently proposed polyethylene terepthalate polymer (PET) or the commonly known Overhead projector (OHP) transparencies (Basu {\it et al.} 2005, 2008) have been shown to have a much higher threshold $\frac{Z}{\beta} \sim 150$. 
A large array of such PET
detectors, as planned in Sandakfu, India, would certainly be a very good system for detection of exotic and rare events.

\section{summary}
QCD transition inside the NSs and possible formation of QSs is one of the most interesting areas of research. 
We have discussed the different facets of
quark star physics along with the associated observational scenarios. Though an observational signature
of phase transition and SQM in nature is still elusive, the study of these phenomena
is extremely important for a better understanding of the physics of strongly interacting matter. Moreover, there are 
many questions remain to be answered before we can make any conclusions.


\bigskip
\bigskip
\noindent {\bf DISCUSSION}

\bigskip
\noindent {\bf G. BISNOVATYI-KOGAN:} For Existence of quark strange stars, their maximal mass should be larger
(or not much smaller) than the maximal mass of neutron stars. Can modern theory make reliable estimate of $M_{max}$
for quark stars?

\bigskip
\noindent {\bf SANJAY K. GHOSH:} Quark stars are self-bound, a characteristics of the underlying QCD physics. So, the 
maximum mass limit estimate for the quark star will carry the uncertainty related to the confinement physics and can be determined only from effective QCD models (Banerjee, Ghosh, Raha 2000).

\bigskip
\noindent {\bf JIM BEALL:} You mentioned that quark stars would be self bound. Does this change the mass limit for these objects?

\bigskip
\noindent{\bf SANJAY K. GHOSH:} Yes. Self bound implies that quark star can be more compact. But then actual estimates will have all the uncertainties of the model parameters. The 
mass-radius relation of quark star comes out to be very different from neutron stars.

\bigskip
\noindent{\bf V. CHECHETKIN:}Direct numerical calculation of baryon-quark phase transition in neutron star shows that the radius of neutron star decrease by 100 meters only.

\bigskip
\noindent{\bf SANJAY K. GHOSH:}The transition is really a two step process - the process of deconfinement and process of strange quark production along with equilibration. So, it is difficult to do a reliable numerical calculation. 
Moreover, the result will depend on the how you are incorporating the confinement.

\bigskip
\noindent{\bf JIM BEALL:} Are there radial modes of oscillation that could produce gravity waves?

\bigskip
\noindent{\bf SANJAY K. GHOSH:}No. Radial modes are basically $l =0$ mode. In dispersion equation 
$l$, gets coupled to the gravity part. So putting  $l=0$  decouples the gravitational radiation.

\bigskip
\noindent{\bf R. BERNABEI:} May you comment on the similar searches already carried out and on the improvements 
of the future ones you are proposing.

\bigskip
\noindent{\bf SANJAY K. GHOSH:}The MACRO as well as SLIM consist of large area passive detectors. So the large area part is certainly not new. On the other hand, strangelets most probably will have high $Z/{\beta}$. So using PET detectors will certainly reduce the large low - $Z/{\beta}$ noise.

\end{document}